# Qualitative Measures of Equity in Small Groups

Ben Archibeque, Kansas State University, barchibeque1@gmail.com
Mary-Bridget Kustusch, DePaul University, kustuschmb@gmail.com
Florian Genz, Universität zu Köln, Florian.Genz@uni-koeln.de
Scott Franklin, Rochester Institute of Technology, svfsps@rit.edu
Eleanor C Sayre, Kansas State University, esayre@gmail.com

**Abstract:** We investigate the utility of two qualitative measures of equity. Our data are videos of groups of first-generation and Deaf or hard-of-hearing students in a pre-matriculation university program designed to help them persist in STEM fields by developing their metacognitive practices. We analyze video data of students in small groups trying to accomplish various tasks. We analyze how groups engage with proposed ideas (inchargeness) and create a space of open sharing (civility). By capturing different aspects of each group, these measures combine to help our understanding of what an equitable group could look like.

## Introduction

Many research-based instructional strategies require small-group work for problem solving or laboratory experiments. We focus on how differences in individuals' behavior within a group affect the participation patterns of all group members. Students' behavior can be harmful or beneficial for their peers, which can hinder or encourage others' willingness to engage with their groupmates. If groups have distributed leadership, all students are more likely to engage with their work (Oliveira, Boz, Broadwell, & Sadler, 2014). However, students who are perceived as less popular or less smart are offered fewer opportunities to participate (Bianchini, 1997) and learn less (Marbach-Ad, Rietschel, Saluja, Carleton, & Haag, 2016).

Esmonde (2009) defines equity "as a fair distribution of opportunities to learn or opportunities to participate." We adapt this definition into two aspects of equity: each group member feels comfortable sharing with their group, and their voice is heard by all group members. The first reflects whether a group allows ideas and opinions to come from all members or only one: if one person is the sole source of ideas, the group is not equitable. The second relates to how members' ideas are taken up within the group: if a group member presents ideas and the rest of the group ignores the ideas, this is also inequitable. In this paper, we draw from Incivility Theory (Cortina et al, 2001) and Positioning Theory (Harré & van Langenhove, 1999) to develop two qualitative measures of equity in small group interactions, and apply them to groups of four undergraduate students modeling climate change in a laboratory setting.

## Theory

Our first aspect of equity is inchargeness, which is rooted in Positioning Theory. Positioning Theory explores how people's communication acts influence and are influenced by what is socially allowed or expected of them (positions) within negotiated storylines (Harré & van Langenhove, 1999). In Positioning Theory, multiple storylines can occur simultaneously, e.g. there could be one storyline between two group members and a different storyline for the whole group that happen concurrently and look very different. This paper focuses on the storyline of the whole group. Inchargeness describes the position of each person in a group and says who controls the flow and topic of conversation (Kustusch et al, 2018). Deitrick (2016) measured positioning between students by counting questions asked and commands given. We analyzed the video itself, not transcripts, and took into account a broader range of communication acts (spoken words, gestures, facial expressions, etc.) to analyze the relative positions of the group in selected segments.

Our second aspect of equity is called civility and is based in Incivility Theory, which focuses on how uncivil acts are ambiguous but are seen as disrespectful or a violation of social norms, including overt behaviors, calling a peer stupid, and covert behaviors, like condescension and social exclusion (Cortina et al., 2001). A group is civil if its members *actively* seek to include their peer's opinions and ideas. Active civility is essential because incivility is frequently experienced but seldom overt (Cortina et al., 2001).

To segment videos, we developed a method which divides a video based on speaking duration changes of group members (Archibeque et al, 2018). These segments vary in size because of that dependence on speaking duration. We analyze the same groups from our previous work.

The research question addressed here is: How do inchargeness and civility inform important aspects of a group's interactions? We will look at both measures for two groups, compare the groups to each other, then compare the measures to one another. Finally, we will discuss future research with these methods.

## Context and Methods

The participants are from a private, doctoral granting technical institute in a two week, pre-matriculation program for first generation (FG) and d/Deaf or hard-of-hearing (DHH) individuals who intend to major in STEM. The program supports their persistence in STEM through metacognitive skill development and community building. Each student identifies with at least one (and often several) underrepresented or marginalized group in higher education. Overall, the race and gender composition of the participants are approximately representative of the general US population in terms of race and gender, with an oversampling of DHH individuals.

This paper analyzes video of two groups (four students in each) developing models for climate change lasting approximately 30 minutes: one group which seemed inequitable and one which seemed equitable based on previous work using quantitative measures of equity (Archibeque et al., 2018); however, the quantitative analysis felt insufficient alone because it did not completely reflect our intuition of the groups. Our intuition was that both groups were inequitable. Here we turn to qualitative measures. In the first group, the students are trying to develop an equation for the amount of carbon dioxide entering the atmosphere annually. In the second group, the students begin by individually journaling "everything they know" about climate change. Then, they were told to reconcile their understandings with their groupmates and develop a model of the atmosphere with the supplies on the table. The groups are from different days in the program, and one student appears in both groups.

To rate inchargeness, multiple raters collaboratively analyzed the communication acts in each segment and rated the inchargeness of each participant during each segment, coming to consensus each time. To operationalize civility, we developed a coding scheme which scores (in)civil behavior across nine axes. For each segment, multiple independent raters select whether that segment was civil, uncivil, and/or neutral on each axis. One axis, for example, is "Group members did not talk over one another more than is socially expected; Group members talked over one another more than is socially expected; Group members did not talk." Because participants might be mixed in their (in)civil behavior, two categories might be selected for the same time segment on the same axis. After rating the video, raters discussed and came to agreement for each segment.

## Results

### Group one

This group consists of Brittany, Jessica, Justin, and Pat. In the first segment, Justin is clearly leading what the group is talking about. When he gets off topic, the others get off topic and vice versa. When he drops a topic, it might be discussed by others, but only briefly before they engage with his new topic. Pat was rated with the next most inchargeness. This is because she is treated as a sage of the group. Whenever there is an idea that someone is unsure of, they ask Pat to make sure it "makes sense." She is also mostly responsible for writing the group's ideas; this is not assigned to her but she does it while Justin is frequently pulling the group toward non-task oriented talk and she engages in off-topic talk less frequently than others. The person with the next most inchargeness in this segment is Brittany. There are several times when Brittany makes a bid to the floor and it is picked up by Justin or Jessica, although not all her bids are picked up. This is true both for task and non-task related bids. Jessica has the least inchargeness. There are several times when she makes bids directly to Justin but is ignored. When they are not directed to Justin, she still has to repeat herself to be acknowledged. There is one point where Justin tells the group to list things which emit carbon dioxide and has his supplies like he will write the ideas down. Then Brittany and Jessica, mostly Brittany, list off ideas but instead of writing down their ideas verbatim, he adapts them (e.g., cars become technology), but when Pat joins in the listing, Justin does not verbalize that he is changing her words like he did for the other two. This showcases Justin's higher positioning overall, Pat's status in the group as someone who is knowledgeable, and Brittany and Jessica's lower positioning overall.

At the end of the episode, an interesting event happens which had potential to change the inchargeness distribution, but did not. After they realized Pat was incorrect and they needed new ideas, Brittany saw an opportunity include Jessica, saying "Why don't we ask Jessica. She is our math person." (Jessica is going to be an applied math major.) Jessica begins speaking, but Justin quickly appears disinterested and writes in his journal, which is a significant departure from how he has been engaging. Shortly after Brittany invites Jessica to speak, the group returns to their original norms: Justin controls the conversation flow, Pat does not interact much but is treated as knowledgeable, and Brittany and Jessica's inchargeness is mostly unchanged. Throughout, Jessica appears to have little say in what the group does or talks about.

This group's speaking time is so heavily dominated by one person that it is difficult to contextualize other people's interactions. Justin is uncivil but the others are, mostly, civil to each other. In the first segment, there were seven uncivil, three civil, and two neutral statements. An exemplar interaction of this segment is between Justin and Jessica. In the span of a few minutes, Justin proposes a few topics to which Brittany and Jessica respond, but only Brittany's response gets feedback from Justin. However, even when he responds, he is clearly looking for her to make specific responses to him and cuts Brittany off frequently.

In the final segment, the event which could have changed the positioning is a two-sided action. On one side, it is civil because Brittany makes an active attempt to include a group member who has been quiet throughout the activity. On the other side, Justin ignores Jessica, which is uncivil. This segment had five civil, five uncivil, and four neutral statements. This segment largely consists of the group sitting in silence. Pat and Justin write separately in their journals and Brittany and Jessica watch them. Eventually Justin asks if what he wrote is okay. It appears as though he shares his journal with Brittany, who holds it so Jessica can see it simultaneously. However, Jessica's inclusion seems to be a result of Brittany's action, not a result of Justin trying to include Jessica.

This group is inequitable according to both qualitative measures. This group has a centralized inchargeness distribution throughout the episode with only slight variations between the people who have smaller amounts of inchargeness, which makes the group very inequitable. The group averages five uncivil, four civil, and three neutral statements, which also makes them inequitable. However, the group also has many neutral statements from the codebook because they speak infrequently. The combination of concentrated inchargeness and incivility show this group as both unwelcoming and silencing. This was the group which we perceived as inequitable, so these measures align with our intuition and quantitative measures.

## Group two

The second group consists of Brock, Herb, Jakob, and Justin. In the first segment, Justin has the most inchargeness but only slightly more than Jakob. When Justin makes bids to the floor, someone in the group always responds to him. He also proposes the most ideas. Jakob's inchargeness is slightly lower because his bids are mostly responses to other bids. It is still high because he openly disagrees with Justin and defends Justin's ideas to Brock and Herb. Herb has the next most inchargeness because he proposes ideas which are occasionally accepted by the group, although he does not speak much. Brock has the least inchargeness because he speaks frequently, but his ideas are ignored by the group. There is one point where he asks "Can I suggest something?" to the group and proposes an idea he had already suggested four times. Only when he asked permission was he allowed to engage with the group. In other cases, he does not ask permission but still has to repeat himself to get his ideas into the group space. Being ignored represents less inchargeness because it shows he has less control over what is talked about and must be allowed into the group. In contrast, Jakob and Justin's ideas are, at worst, acknowledged by the other.

Near the end of the first segment, Jakob begins to explain an idea and indicates he wants to monologue. Before he begins, the teacher comes over to ask what the group is doing. Brock says they are still deciding. Jakob responds with "that is what I am doing" and proceeds to elaborate his thoughts to the instructor. In this case, he uses the instructor's question to state his ideas with minimal interruption and take more inchargeness. When the second segment begins, Jakob is a few seconds into his lengthy explanation. In this segment, Jakob and Justin have equal amounts of inchargeness. It is clear that Jakob's push to be heard in the group is received. Justin still has as much inchargeness, however, because even though he is not proposing ideas, he is clearly open to providing Jakob feedback. Herb's inchargeness has shifted down, below Brock, in this segment. There is little to no time allotted here for him to engage with his group; the other group members dominate discussion. Brock's role has not changed because he is still making bids which are mostly ignored.

In terms of civility, in the first segment, Brock is ignored by his peers as he repeatedly brings up an idea but is only acknowledged after several bids; he literally asks to make a suggestion at one point. Another example of the group's civility is when Justin and Jakob are simultaneously talking and clearly want to say something about what to do next. They talk over each other until Jakob begins to make a 'baap' noise any time Justin tries to say something. Eventually Justin stops talking and Jakob begins to explain what he wanted to. In this segment there are four civil, seven uncivil and no neutral statements.

The second segment is mostly Jakob explaining to the instructor. This group is kind about Jakob's explanation, mostly does not interrupt him, and occasionally contributes to the discussion instead of holding fast to their opinions without reason. This segment has five civil, four uncivil, and five neutral statements.

We expected this group to be more equitable based on the quantitative measures but inequitable according to our intuition. According to these two qualitative measures, it is inequitable. Inchargeness is distributed between two people, which means it is more equitable than the first group, but is clearly not evenly distributed, which means it is still inequitable. Brock and Herb clearly do not have a voice in the group. This group averages five civil, five uncivil, and two neutral statements, which is slightly more civil than Group One but has the same amount of incivility. The civility of the group was difficult to reconcile because they were inconsistent in their actions. Brock was both ignored and acknowledged by the same people in a short period of time; they all talked over one another but, some of this time, they were engaging with one another's ideas.

## Discussion and conclusion

The first measure of equity was the distribution of inchargeness. In both groups, there was a concentration of inchargeness, as opposed to an imagined, highly equitable group, which has an even distribution of inchargeness. This measure shows who holds the power of the group. In G1, Justin had the most inchargeness, which put more weight on his actions. Thus, when Jessica was encouraged by Brittany to have a voice in the group and he ignored her, his actions had a bigger impact than Brittany's, so Jessica quickly stopped speaking. Inchargeness could be further developed and verified as an effective tool to measure the positioning of students in contrast to Dietrick's "Commands" and "Questions" operationalization. In both cases, the concentration of inchargeness still negatively impacted those who did not have it. Those with less inchargeness spoke less and engaged less, which could mean they will learn less (Marbach-Ad et al., 2016) from this program.

The second measure was the civility of the group. Both groups were similarly uncivil. In G1, it was due to some individuals being civil and a different individual being uncivil. For G2, this was due to a group member's persistence; if Brock was ignored, he repeated himself until he was acknowledged. Getting ignored is uncivil, but later getting acknowledged is civil. We choose to analyze on the group level because if a group member is being silenced, it matters that the group allows that incivility to be a part of the discourse. This method is helpfully describes how comfortable a group is for all its members. A solely civil group is probably comfortable for all members. Both groups we analyzed average around fifty percent civil (ignoring neutral statements).

When we look at how uncivility was moderated by inchargeness, Justin in G1 left little room for others to have authority so any civil motion (without his support) was insignificant. In G2, it is possible that the division of power made Brock feel comfortable repeating himself.

These groups were chosen because they seemed dissimilar quantitatively, but our intuition about them was that they were more alike in terms of inequity, which was validated by the qualitative measures described here. We hope to see a continued use of comprehensive measures of equity so that classroom interactions can be understood more holistically. We developed these methods with the hope for more rapid, possibly *in vivo*, analysis of any type of group (various sizes, different settings, etc.) and that individuals' group experiences in the program can be related to their persistence in it. One weakness of this study is that we selected these data because they stood out. Work should be done to see if these methods effectively characterize groups which are more typical.

## References

Archibeque, B., Genz, F., Franklin, M., Franklin, S., & Sayre, E. (2018). Quantitative Measures of Equity in Small Groups. In *Physics Education Research Conference*. Cinncinati, OH.

Bianchini, J. A. (1997). Where knowledge construction, equity, and context intersect: Student learning of science in small groups. *Journal of Research in Science Teaching*, *34*(10), 1039–1065.

Cortina, L. M., Magley, V. J., Williams, J. H., & Langhout, R. D. (2001). Incivility in the workplace: incidence and impact. *Journal of Occupational Health Psychology*, *6*(1), 64–80.

Deitrick, E., Shapiro, R. B., & Gravel, B. (2016). How do wo Assess Equity in Programming Pairs? In C.-K. Looi, J. Polman, U. Cress, & P. Reimann (Eds.) (Vol. 1, pp. 370–377). Presented at ICLS, Singapore: International Society of the Learning Sciences, Inc.

Esmonde, I. (2009). Ideas and Identities: Supporting Equity in Cooperative Mathematics Learning. *Review of Educational Research*, *79*(2), 1008–1043.

Harré, R., & van Langenhove, L. (1999). Introducing Positioning Theory. In R. Harré & L. van Langenhove (Eds.), *Positioning Theory: Moral Contexts of International Action* (1st ed., pp. 14–31). Malden, Massachusetts: Blackwell Publishers Inc.

Kustusch, M.B., Sayre, E.C., & Franklin, S.V. (2018) Identifying shifts in agency by analyzing authority in classroom group discussion. In C.-K. Looi, J. Polman, U. Cress, & P. Reimann (Eds.). Presented at ICLS, London: International Society of the Learning Sciences, Inc.

Marbach-Ad, G., Rietschel, C. H., Saluja, N., Carleton, K. L., & Haag, E. S. (2016). The Use of Group Activities in Introductory Biology Supports Learning Gains and Uniquely Benefits High-Achieving Students. *Journal of Microbiology & Biology Education*, *17*(3), 360–369.

## Acknowledgements

The authors are deeply grateful to the Equity Working Group of the PEER-Rochester field school and research group. Portions of this research were funded by NSF grants; the Developing Scholars program and Physics Department at KSU; CASTLE at RIT; and the German Federal Ministry of Education and Research. The authors are responsible for the content of this publication.